\documentstyle[aps,twocolumn,psfig,epsf]{revtex}

\begin{document}
\draft
\title{\large \bf RELATIVISTIC INSTANT--FORM APPROACH TO
THE STRUCTURE OF
TWO-BODY COMPOSITE SYSTEMS. NONZERO SPIN}

\author{A.~F.~Krutov\thanks{Electronic address:
krutov@ssu.samara.ru}}

\address {\it Samara  State University, RU-443011, Samara,
Russia}

\author{V.~E.~Troitsky\thanks{Electronic address:
troitsky@theory.sinp.msu.ru}}
\address {\it D.~V.~Skobelltsyn Institute of Nuclear Physics ,
Moscow State University, RU-119899, Moscow, Russia}

\date{April, 2003}
\maketitle
\begin{abstract}
The relativistic approach to electroweak properties
of two-particle composite systems developed in  Ref.\cite{KrT02}
is generalized here to the case of nonzero spin.
In developed technique the parametrization of matrix elements of electroweak
current operators in terms of form factors is a realization of the
Wigner--Eckart theorem on the Poincar\'e group and form factors
are reduced matrix elements. The $\rho$ meson charge form factor
is calculated as an example.\end{abstract}

PACS number(s): 13.40.--f, 11.30.Cp
\narrowtext

A new relativistic approach to electroweak properties of
composite systems has been proposed in our recent paper
\cite{KrT02}. The approach is based on the use of the instant form (IF)
of relativistic Hamiltonian dynamics (RHD).
The detailed description of RHD can be found in the review
\cite{KeP91}.
Some other
references as well as some basic equations of RHD
approach are given in  Ref.\cite{KrT02}.

Now our aim is to generalize the approach
to composite systems of two particles of
spin 1/2 with nonzero values of total angular momentum,
total orbital momentum and total spin.  The main problem is
a construction of electromagnetic current operator satisfying
standard conditions
(see, e.g., Refs.\cite{KrT02,KrT02hep}).

The basic point of our approach
\cite{KrT02} to the construction of the electromagnetic current
operator is the general method of relativistic invariant parameterization of
local operator matrix elements proposed as long ago as in
1963 by Cheshkov and Shirokov~\cite{ChS63}.
This canonical parametrization of local
operators matrix elements was generalized to the case of composite systems
of free particles in Refs.~\cite{TrS69,KoT72}.
This parametrization is a
realization of the Wigner--Eckart theorem for the Poincar\'e
group and so it enables one for given matrix element of arbitrary
tensor dimension to separate the reduced
matrix elements (form factors) that are invariant
under the Poincar\'e group.

Physical approximations that we use in our approach are
formulated in terms of reduced matrix elements, for example, the
well known relativistic impulse approximation.
In our method this
approximation does not violate the standard conditions
for the current.

In the present paper we propose a general formalism for the
operators diagonal in the total angular momentum.
The details of calculations can be found in  \cite{KrT02hep}.

Let us consider the operator
$j_\mu = j_\mu(0)$ that describes a transition between two states of a
composite two-- constituent system.
Let us neglect temporarily for simplicity the
conditions of self--adjointness, conservation law and parity
conservation.
The Wigner--Eckart
decomposition of the matrix element has the form \cite{ChS63}:
\begin{eqnarray}
\langle\,\vec p_c,\,m_{Jc}\,|j_\mu|\,\vec p_c\,',m'_{Jc}\,\rangle
= \langle\,m_{Jc}|D^{J_c}(p_c,\,p_c')\nonumber\\
\times\left[\,F^c_1\,K'_\mu + F^c_2\,\Gamma_\mu (p_c')  \right.
\left. + F^c_3\,R_\mu + F^c_4\,K_\mu\right]|m'_{Jc}\rangle\;,
\label{<|jc|>=F_is}\\
F^c_i = \sum _{n=0}^{2J_c}\,f^c_{in}(Q^2)(ip_{c\mu}\Gamma^\mu(p_c'))^n\;.
\label{Fic}
\end{eqnarray}
Here $K_\mu=(p_c - p_c')_\mu=q_\mu, K'_\mu=(p_c + p_c')_\mu, 
R_\mu=\epsilon _{\mu\nu\lambda\rho}\,p_c^\nu\,p_c'\,^\lambda\Gamma^\rho(p_c'); 
(p_c - p_c')^2=-\,Q^2,\,p_{c}^2=p'_{c}\,^2=M_c^2,\,M_c,\,J_c$ are the
mass and spin of the composite particle, $m_{Jc}$ is spin
projection, $\Gamma^\rho (p_c')$ is the spin
four--vector defined with the use of the Pauli--Lubansky vector
\cite{KrT02}, $f^c_{in}$ are reduced matrix elements,
$\epsilon _{\mu\nu\lambda\rho}$ is a
completely antisymmetric pseudo-tensor in four dimensional
space-time with $\epsilon _{0\,1\,2\,3}= -1$.

In the frame of RHD the form factors of composite systems
$f^c_{in}$ are to be expressed in terms of RHD wave
functions and constituents form factors.

In RHD a state of two particle interacting system is described
by a vector in the direct product of two one--particle Hilbert
spaces (see, e.g., Ref.~\cite{KrT02}). So, the matrix element in
RHD can be decomposed in the basis \cite{KrT02}
\begin{equation}
|\,\vec P,\;\sqrt {s},\;J,\;l,\;S,\;m_J\,\rangle\;.
\label{bas-cm}
\end{equation}
Here $P_\mu = (p_1 +p_2)_\mu$, $P^2_\mu = s$, $\sqrt {s}$
is the invariant mass of the two-particle system,
$l$ is the orbital angular momentum in the center--of--mass
frame (c.m.), $S$ is the total spin in the c.m., $J$ is the
total angular momentum with the projection $m_J$.
\begin{eqnarray}
\langle\vec p_c,m_{Já}|j_\mu|\vec p_c\,',m'_{Já}\rangle =
\sum\int\frac{d\vec P\,d\vec P\,'}{N_{CG}\,N_{CG}'}\,
d\sqrt{s}\,d\sqrt{s'}\,\nonumber\\
\times\langle\,\vec p_c,m_{Jc}|\vec P,\sqrt{s},J,l,S,m_J\rangle\nonumber\\
\times\langle\vec P,\sqrt{s},J,l,S,m_J|j_\mu|
\vec P\,',\sqrt{s'},J',l',S',m_{J'}\rangle\nonumber\\
\times\langle
\vec P\,',\sqrt{s'},J',l',S',m_{J'}|\vec p_c\,',m'_{Jc}\rangle\;.
\label{j=int}
\end{eqnarray}
Here the sum is over variables $J$,$J'$,$l$,$l'$,$S$,$S'$,$m_J$,$m_{J'}$, and
$\langle\vec P\,',\sqrt{s'},J',l',S',m_{J'}|
\vec p_á\,',m'_{Jc}\rangle$ is the wave function in the sense of IF RHD.
\begin{eqnarray}
\langle\vec P\,\,,\sqrt{s}\,,J\,,l\,,S\,,m_{J}|\,\vec p_c\,,m_{J_c}\rangle
\nonumber\\
= N_c\,\delta (\vec P\, - \vec p_c)\delta_{J_cJ}\delta_{m_{J_c}m_{J}}
\,\varphi^{J_c}_{lS}(k)\;.
\label{wf}
\end{eqnarray}
Here $k = {\sqrt{\lambda(s\,,\,M_1^2\,,\,M_2^2)}}/{(2\sqrt{s})},$
$M_1\,,\,M_2$ are masses of constituents,
$\lambda (a,b,c) = a^2 + b^2 + c^2 - 2(ab + bc + ac)$.

The RHD wave function of constituents relative motion with
fixed total angular momentum is defined as
\begin{equation}
\varphi^{J_c}_{lS}(k(s)) =\sqrt{\sqrt{s}(1 - \eta^2/s^2)}\,u_{lS}(k)\,k\;,
\label{phi(s)}
\end{equation}
and is normalized by the condition
\begin{equation}
\sum_{lS}\int\,u_{lS}^2(k)\,k^2\,dk = 1\;.
\label{norm}
\end{equation}
Here $\eta = M_1^2 - M_2^2\,$,$\,u_{lS}(k)$  is a model wave
function.

The main difficulty arising in this case is the following.
In the expression (\ref{<|jc|>=F_is}) we were dealing with the
parametrization of local operator matrix elements in the case when the
transformations of the state vectors and of the operators were
defined by  one and the same representation of the quantum
mechanical Poincar\'e group.

A different situation takes place in the case of the
matrix element in the r.h.s. of
Eq.~(\ref{j=int}). The operator describes the system of two
interacting particles and transforms following the
representation with Lorentz boosts generators depending on the
interaction \cite{KrT02}.
The state vectors physically describe the system
of two free particles and present the basis of a representation
with interaction--independent generators. So, the Wigner--Eckart
decomposition can not be applied
directly to the matrix element in the integrand in the r.h.s. of
Eq.~(\ref{j=int}). This is caused by the fact that it is
impossible to construct 4--vectors describing the matrix element
transformation properties under the action of Lorentz boosts
from the variables entering the state vectors (contrary to the
case of, e.g., Eq.~(\ref{<|jc|>=F_is})).
In fact, the possibility of matrix
element representation in the form (\ref{<|jc|>=F_is})
is based on the following fact. Let us act by Lorentz
transformation on the operator $\hat U^{-1}(\Lambda)j^\mu\hat U(\Lambda) =
\tilde j^\mu$.
We obtain the following chain of equalities:
\begin{eqnarray}
\langle \vec p_c,m_{Jc}|\tilde j^\mu|\vec p_c\,',m'_{Jc}\rangle\nonumber\\
= \langle \vec p_c,m_{Jc}|\hat U^{-1}(\Lambda)j^\mu\hat U(\Lambda)
|\vec p_c\,',m'_{Jc}\rangle\nonumber\\
= \sum_{\tilde m_{Jc},\tilde m'_{Jc}}
\langle\,m_{Jc}|[D^{J_c}(R_\Lambda)]^{-1}|\,\tilde m_{Jc}\rangle\nonumber\\
\times\langle \Lambda \vec p_c,\tilde m_{Jc}|j^\mu|
\Lambda \vec p_c\,',\tilde m'_{Jc}\rangle
\langle\,\tilde m'_{Jc}|D^{J_c}(R_\Lambda)|\,m'_{Jc}\rangle.
\label{tj=LpjLp}
\end{eqnarray}
Here $D^{J_c}(R_\Lambda)$
is rotation matrix realizing the angular momentum
transformation under the action of Lorentz transformations.
The equalities (\ref{tj=LpjLp})
show that the transformation properties of the current as a
4--vector can be described using the 4--vectors
of the initial and the final states. This means that
the canonical parameterization \cite{ChS63} is the realization
of the Wigner--Eckart theorem on the Poincar\'e group.

In the case of the current matrix element in the r.h.s. of
Eq.~(\ref{j=int}) the relations
(\ref{tj=LpjLp}) are not valid and direct application
of the Wigner--Eckart theorem is impossible.

However, it can be shown that for the matrix element in
Eq.~(\ref{j=int}) considered as a generalized function
(distribution), that is
considered as an object having sense only under integrals and
sums in
Eq.~(\ref{j=int}). So, the equality (\ref{tj=LpjLp}) is valid in
the weak sense.

Let us consider the matrix element in question as a regular
Lorentz covariant generalized function (see, e.g.,
Ref.~\cite{BoL87}). Using Eq.~(\ref{wf}) let us rewrite
Eq.~(\ref{j=int}) in the following form:
\begin{eqnarray}
\langle\vec p_c,m_{Jc}|j_\mu|\vec p_c\,',m'_{Jc}\rangle\nonumber\\
= \sum_{l,l',S,S'}\int\,{\cal N}\,
d\sqrt{s}\,d\sqrt{s'}\,\varphi^{Jc}_{lS}(s)\varphi^{Jc}_{l'S'}(s')
\nonumber\\
\times\langle\vec p_c,\sqrt{s},J_c,l,S,m_{Jc}|j_\mu|
\vec p_c\,',\sqrt{s'},J_c,l',S',m'_{Jc}\rangle\;.
\label{j=int ds}
\end{eqnarray}
Here it is taken into account that the current operator $j_\mu$
is diagonal in total angular momentum of the composite system,
${\cal N} = {N_c\,N'_c}/{N_{CG}\,N_{CG}'}$.

Let us make use of the fact that the set of the states
(\ref{bas-cm}) is complete:
\begin{eqnarray}
\hat I = \sum\int\,\frac{d\vec P}{N_{CG}}\,d\sqrt{s}\nonumber\\
\times|\vec P,\sqrt{s},J,l,S,m_{J}\rangle
\langle\vec P,\sqrt{s},J,l,S,m_{J}|\;.
\label{I=compl}
\end{eqnarray}
Here the sum is over the discrete variables of the basis
(\ref{bas-cm}).

Under the integral the matrix element of the transformed current
satisfies the following equalities
((\ref{wf}) and  (\ref{I=compl}) are taken into account):
\begin{eqnarray}
\sum\int\,{\cal N}\,d\sqrt{s}\,d\sqrt{s'}\,
\varphi^{Jc}_{lS}(s)\varphi^{Jc}_{l'S'}(s')\nonumber\\
\times
\langle\vec p_c,\sqrt{s},J_c,l,S,m_{Jc}|
\hat U^{-1}(\Lambda)j_\mu\hat U(\Lambda)\nonumber\\
\times|\vec p_c\,',\sqrt{s'},J_c,l',S',m'_{Jc}\rangle\nonumber\\
= \sum\int\,{\cal N}\,
d\sqrt{s}\,d\sqrt{s'}\,\varphi^{Jc}_{lS}(s)\varphi^{Jc}_{l'S'}(s')\nonumber\\
\times\sum_{\tilde m_{Jc},\tilde m'_{Jc}}
\langle\,m_{Jc}|[D^{J_c}(R_\Lambda)]^{-1}|\,\tilde m_{Jc}\rangle\nonumber\\
\times\langle\Lambda\vec p_c,\sqrt{s},J_c,l,S,\tilde m_{Jc}|j_\mu
|\Lambda\vec p_c\,',\sqrt{s'},J_c,l',S',\tilde m'_{Jc}\rangle\nonumber\\
\times\langle\,\tilde m'_{Jc}|D^{J_c}(R_\Lambda)|\,m'_{Jc}\rangle\;.
\label{j=int3}
\end{eqnarray}

It is easy to see that under the integral the current matrix
element satisfies the equalities analogous to
Eq.~(\ref{tj=LpjLp}), so now it is possible to use the
parameterization under the
integral, that is to use the Wigner--Eckart theorem in the weak
sense. The r.h.s. of Eq.~(\ref{j=int ds}) can be written as a
functional on the space of test functions of the form (see
Eq.~(\ref{phi(s)}), too))
$
\psi^{ll'SS'}(s\,,\,s') = u_{lS}(k(s))\,u_{l'S'}(k(s'))\;,
$
and Eq.~(\ref{j=int ds}) can be
rewritten as a functional in {\bf R}$^2$ with variables $(s,s')$:
\begin{eqnarray}
\langle\vec p_c\,,m_{Jc}|j_\mu(0)|\vec p_c\,'\,,m'_{Jc}\rangle\nonumber\\
= \sum_{l,l',S,S'}\int\,d\mu(s,s'){\cal N}\,
\psi^{ll'SS'}(s,s')\nonumber\\
\times\langle\vec p_c,\sqrt{s},J_c,l,S,m_{Jc}|j_\mu|
\vec p_c\,',\sqrt{s'},J_c,l',S',m'_{Jc}\rangle\;.
\label{j=int dmu}
\end{eqnarray}
Here the measure
is chosen with the account of the
relativistic density of states, subject to the normalization
(\ref{phi(s)}), (\ref{norm}):
\begin{eqnarray}
d\mu(s,s') = 16\,\theta(s - (M_1+M_2)^2)\,\theta(s' - (M_1+M_2)^2)\nonumber\\
\times\sqrt{\sqrt{s}(1 - \eta^2/s^2)\sqrt{s'}(1 - \eta^2/{s'\,}^2)}\,
d\mu(s)\,d\mu(s')\;.
\label{dmuM1M2}
\end{eqnarray}
Here $d\mu(s)=({1}/{4})\,k\,d\sqrt{s}$.

The sums over discrete invariant variables can be transformed
into integrals by introducing the adequate delta--functions.
The obtained expressions are functionals
in {\bf R}$^6$.

The functional in the r.h.s. of
Eq.~(\ref{j=int dmu}) defines a Lorentz covariant generalized
function, generated by the current operator matrix element.

Taking into account
Eq.~(\ref{j=int3}) we decompose the matrix element in the r.h.s.
of Eq.~(\ref{j=int dmu}) into the set of linearly independent
scalars entering the r.h.s. of Eq.(\ref{<|jc|>=F_is}):
\begin{eqnarray}
{\cal N}\langle\vec p_c,\sqrt{s},J_c,l,S,m_{Jc}|j_\mu|\vec
p_c\,',\sqrt{s'},J_c,l',S',m'_{Jc}\rangle\nonumber\\
= \langle\,m_{Jc}|D^{J_c}(p_c,\,p_c')
\sum_{n=0}^{2J_c}(ip_{c\mu}\Gamma^\mu(p_c'))^n\nonumber\\
\times{\cal A}^{ll'SS'}_{n\mu}(s,Q^2,s')|m'_{Jc}\rangle\;.
\label{<|jc|>=cA}
\end{eqnarray}
Here ${\cal A}^{ll'SS'}_{n\mu}(s,Q^2,s')$ is a
Lorentz covariant generalized function.

Making use of Eq.~(\ref{<|jc|>=cA}) and comparing
the r.h.s. of
Eq.~(\ref{<|jc|>=F_is}) with Eq.~(\ref{j=int dmu}) we obtain:
\begin{eqnarray}
\sum_{l,l',S,S'}\int\,d\mu(s,s')\,
\psi^{ll'SS'}(s,s')\nonumber\\
\times\langle\,m_{Jc}|{\cal A}^{ll'SS'}_{n\mu}(s,Q^2,s')\,|m'_{Jc}\rangle
\nonumber\\
= \langle\,m_{Jc}|\left[\,f^c_{1n}\,K'_\mu + f^c_{2n}\,\Gamma_\mu (p'_c)
\right.\nonumber\\
\left.
+ f^c_{3n}\,R_\mu + f^c_{4n}\,K_\mu\right]|m'_{Jc}\rangle\;.
\label{c=c}
\end{eqnarray}

All the form factors in the r.h.s. of
Eq.~(\ref{c=c}) are nonzero if the generalized function ${\cal
A}$ contains parts that are diagonal (${\cal A}_1$) and
non-diagonal (${\cal A}_2$) in $m_{Jc}\,,\,m'_{Jc}$
For the diagonal part we have from
Eq.~(\ref{c=c}):
\begin{eqnarray}
\sum_{l,l'S,S'}\int\,d\mu(s,s')\,
\psi^{ll'SS'}(s,s')\nonumber\\
\langle\,m_{Jc}|{\cal A}^{ll'SS'}_{1n\mu}(s,Q^2,s')\,|m_{Jc}\rangle\nonumber\\
= \langle\,m_{Jc}|\left[\,f^c_{1n}[\psi]\,K'_\mu +
f^c_{4n}[\psi]\,K_\mu\right]|m_{Jc}\rangle\;.
\label{cd=cd}
\end{eqnarray}
The notation $f^c_{in}[\psi]$ in the r.h.s. emphasizes the fact
that form factors of composite systems are
functionals on the wave functions of the intrinsic motion
and so, on the test functions.

Let the equality
(\ref{cd=cd}) be valid for any test function
$\psi^{ll'SS'}(s,s')$.
When the test functions (the intrinsic motion wave functions)
are changed the vectors in the r.h.s. are not changed because
according to the essence of the parametrization
(\ref{<|jc|>=F_is}) they do not depend on the model for the
particle intrinsic structure. So, when the test functions are
varied the vector of the r.h.s. of Eq.~(\ref{cd=cd})
remains in the hyperplane defined by the vectors
$K_\mu\,,\,K'_\mu$.

When test functions are varied arbitrarily the vector in
l.h.s. of Eq.~(\ref{cd=cd}) can take, in
general, an arbitrary direction. So, the requirement of the
validity of Eq.~(\ref{cd=cd}) in the whole space of our test
functions is that the l.h.s. generalized function have
the form:
\begin{eqnarray}
{\cal A}^{ll'SS'}_{1n\mu}(s,Q^2,s') = K'_\mu\,G^{ll'SS'}_{1n}(s,Q^2,s')
\nonumber\\
+ K_\mu\,G^{ll'SS'}_{4n}(s,Q^2,s')\;.
\label{cA1}
\end{eqnarray}
Here $G^{ll'SS'}_{in}(s,Q^2,s')\;,\; i=1,4$
are Lorentz invariant generalized functions. Substituting
Eq.~(\ref{cA1}) in Eq.~(\ref{cd=cd}) and taking into account
Eqs.~(\ref{phi(s)}) and (\ref{dmuM1M2})
we obtain the following integral representations:
\begin{eqnarray}
f^c_{in}(Q^2)
= \sum_{l,l',S,S'}\int\,d\sqrt{s}\,d\sqrt{s'}\,
\varphi^{Jc}_{lS}(s)\,\varphi^{Jc}_{l'S'}(s')
\nonumber\\
\times\,G^{ll'SS'}_{in}(s,Q^2,s')\;
\label{intrep}
\end{eqnarray}
for $i = 1,4$.
In the case of matrix element in
Eq.~(\ref{c=c}) non-diagonal in $m_{Jc}\,,\,m'_{Jc}$ we can
proceed in an analogous way and obtain an analogous integral
representations for $f^c_{in}(Q^2)\;,\;i=2,3$.

So, the matrix element in the r.h.s. of
Eq.~(\ref{j=int dmu}) considered as Lorentz covariant
generalized function can be written as the following
decomposition of the type of Wigner--Eckart decomposition:
\begin{eqnarray}
\langle\vec p_c,\sqrt{s},J_c,l,S,m_{Jc}|j_\mu|
\vec p_c\,',\sqrt{s'},J_c,l',S',m'_{Jc}\rangle\nonumber\\
= \frac{1}{{\cal N}}\,
\langle\,m_{Jc}|D^{J_c}(p_c,\,p_c')\left[\,{\cal F}_{1}\,K'_\mu +
{\cal F}_{2}\,\Gamma^\mu (p'_c) \right.\nonumber\\
\left.
+ {\cal F}_{3}\,R_\mu + {\cal F}_{4}\,K_\mu\right]|m'_{Jc}\rangle\;.
\label{fin}\\
{\cal F}_i = \sum _{n=0}^{2J_c}\,G^{ll'SS'}_{in}(s,Q^2,s')
(ip_{c\mu}\Gamma^\mu(p_c'))^n\;.
\label{cfic}
\end{eqnarray}

\begin{figure}[htbp]\vspace*{-0.5cm}
\epsfxsize=0.9\textwidth
\centerline{\psfig{figure=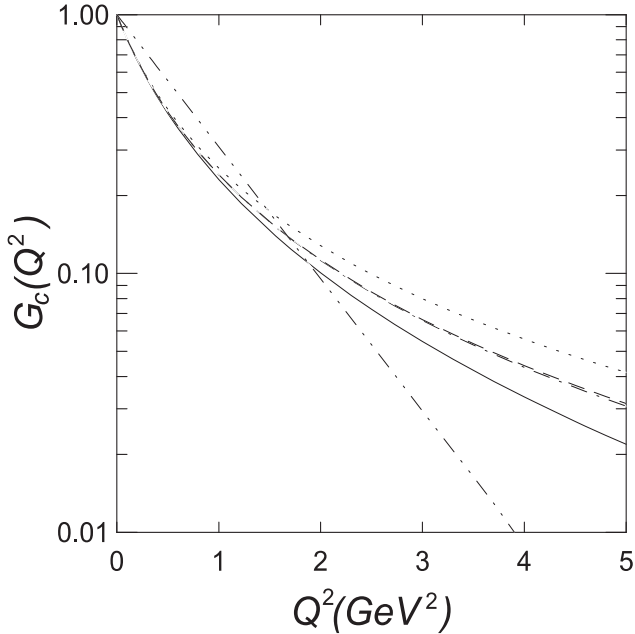,height=8.6cm,width=8.6cm}}
\vspace{0.1cm}
\caption{The results of the calculations of the
$\rho$-meson charge form factor with different model wave
functions \protect\cite{KrT02,KrT02hep}.
The solid line represents the relativistic calculation with the wave
function of harmonic oscillator,
the dashed line -- with the power--law wave function
for $n =$ 3, dash--dot--line -- with the wave function with linear
confinement, dotted
line -- with the power--law wave function for $n =$ 2,
dot--dot--dash-line --
the non-relativistic calculation with the wave
function of harmonic oscillator.
The wave functions parameters are obtained from the fitting of
$\rho$ -- meson MSR. The sum of quark anomalous magnetic moments
is taken as
$\kappa_u + \kappa_{\bar d}$ = 0.09 in natural units. The quark mass is
$M$ = 0.25 GeV.}
\label{fig:1}
\end{figure}
\vspace{-0.2cm}

In Eqs.~(\ref{fin}), and (\ref{cfic})
the form factors $G^{ll'SS'}_{in}(s,Q^2,s')$ contain all the information
about the physics
of the transition described by the operator
$j_\mu$. They are connected with the composite particle form
factors (\ref{<|jc|>=F_is}), and (\ref{Fic})
through Eq.~(\ref{intrep}). In particular, physical
approximations are formulated in our approach in terms of form
factors $G^{ll'SS'}_{in}(s,Q^2,s')$ (see Ref.~\cite{KrT02} for
details).  The matrix element transformation properties are
given by the 4--vectors in the r.h.s. of Eq.~(\ref{fin}).

It is worth to emphasize that it is necessary to consider the
composite system form factors as the functionals generated
by the Lorentz invariant generalized functions
$G^{ll'SS'}_{in}(s,Q^2,s')$.

Now let us impose the conditions of self--adjointness, conservation law
and parity conservation
on the matrix
elements in Eqs.~(\ref{<|jc|>=F_is}), and (\ref{fin}). The
r.h.s.  of equalities (\ref{<|jc|>=F_is}) and  (\ref{fin})
contain the same 4--vectors and the same sets of
Lorentz scalars (\ref{Fic}) and (\ref{cfic}), so, to take into
account the additional conditions it is necessary to redefine
these 4--vectors and functions $G^{ll'SS'}_{in}(s,Q^2,s')$.
For example, the
conservation law gives $F_4 ^c=0$ and ${\cal F}_{4}$=0.

Let us write the parameterization
(\ref{fin}), (\ref{cfic}) for the particular case of composite particle
electromagnetic current with quantum numbers
$J=J'=S=S'=1$, which is realized, for example, in the case of
deuteron. Separating the quadrupole form factor
and using
Eqs.~(\ref{fin}), and (\ref{cfic}) we obtain the following form:
\begin{eqnarray}
\langle\vec p_c,\sqrt{s},J_c,l,S,m_{Jc}|j_\mu|
\vec p_c\,',\sqrt{s'},J_c,l',S',m'_{Jc}\rangle\nonumber\\
= \frac{1}{{\cal N}}\langle\,m_{Jc}|\,D^{1}(p_c\,,p'_c)\,
\left[
\tilde{\cal F}_1\,K'_\mu
+ \frac{i}{M_c}\tilde{\cal F}_3\,R_\mu\right]|m'_{Jc}\rangle\;.
\label{J=1}\\
\tilde{\cal F}_1 = \tilde G^{ll'}_{10}(s,Q^2,s') +
\tilde G^{ll'}_{12}(s,Q^2,s')\left\{[i{p_c}_\nu\,\Gamma^\nu(p'_c)]^2
\right.\nonumber\\
\left. - \frac{1}{3}\,\hbox{Sp}[i{p_c}_\nu\,\Gamma^\nu(p'_c)]^2\right\}
\frac{2}{\hbox{Sp}[{p_c}_\nu\,\Gamma^\nu(p'_c)]^2}\;,\nonumber\\
\tilde{\cal F}_3 = \tilde G^{ll'}_{30}(s,Q^2,s')\;.
\label{FJ=1}
\end{eqnarray}

We have taken into account that the equation
$\tilde G^{ll'}_{21}(s,Q^2,s') =$ 0 is valid in weak
sense.  The parametrization (\ref{<|jc|>=F_is}), (\ref{Fic}) takes the
form  (\ref{J=1}), (\ref{FJ=1}) with
$\tilde G^{ll'}_{in}(s,Q^2,s')\to\tilde f^c_{in}(Q^2)$.
It is easy to see that
for the redefined form factors the equality (\ref{intrep})
remains valid.
Form factors $\tilde f^c_{in}(Q^2)$ are connected with
charge, quadrupole and magnetic Sachs form factors:
$
G_C(Q^2)=\tilde f^c_{10}(Q^2),
G_Q(Q^2)=({2\,M_c^2}/{Q^2})\tilde f^c_{12}(Q^2),
G_M(Q^2)=-M_c\tilde f^c_{30}(Q^2).
$
The modified impulse approximation (MIA) can be
formulated in terms of form factors
$\tilde G^{ll'}_{iq}(s,Q^2,s')$.
The physical meaning of this approximation is considered in
detail in Ref.~\cite{KrT02,KrT02hep}.

The results of calculations for the $\rho$ -- meson charge form
factor in MIA  ($l=l'=0$) are represented in Fig.1.

The authors thank S.Simula for the discussion of
RHD problems (see Ref.\cite{MeS02}).

This work was supported in part by the Program "Russian
Universities--Basic Researches" (Grant No. 02.01.013)
and Ministry of Education of Russia (Grant No. E02--3.1--34).

\end{document}